\documentclass[epj]{svjour}

\usepackage{graphicx}
\usepackage{fancyhdr}
\usepackage{amssymb,amsmath}
\def\makeheadbox{{
 }}
\def\mytitle{My title} 
\def\myauthors{My name}  
\def\mytype{My type of session}
\def\mysession{My session}

\def\mytitle{Indirect search for Dark Matter with the ANTARES neutrino telescope} 
\def\myauthors{Gordon Lim}    
\def\mytype{Contributed Talk}    
\def\mysession{Cosmology and Astrophysics}

\begin{document}
\title{Indirect search for Dark Matter with the ANTARES neutrino telescope}

\author{G.M.A. Lim \thanks{\emph{Email:} gordonl@nikhef.nl} on behalf of the ANTARES collaboration}

\institute{NIKHEF, PO Box 41882, 1009 DB Amsterdam, The Netherlands}

\date{}
\abstract{
Relic neutralinos produced after the Big Bang are favoured candidates for Dark Matter. They can accumulate at the centre of massive celestial objects like our Sun. Their annihilation can result in a high-energy neutrino flux that could be detectable as a localised emission with earth-based neutrino telescopes like ANTARES. In this paper a brief overview of the prospects of the indirect search for Dark Matter particles with the ANTARES detector will be given. The analysis method and expected performance for the detection of the expected neutrinos will be discussed.
} 
\maketitle

\section{Introduction}
\label{intro}
It is now a well-established fact that according to our present theory of gravity, 85\%~of the matter content of our universe is missing. Observational evidence for this discrepancy ranges from macroscopic to microscopic scales, e.g. gravitational lensing in galaxy clusters, galactic rotation curves and fluctuations measured in the Cosmic Microwave Background. This has resulted in the hypothesised existence of a new type of matter called Dark Matter. Particle physics provides a well-motivated explanation for this hypothesis: The existence of (until now unobserved) massive weakly interacting particles (WIMPs). A favorite amongst the several WIMP candidates is the neutralino, the lightest particle predicted by Supersymmetry, itself a well-motivated extension of the Standard Model. 

If Supersymmetry is indeed realised in Nature, Supersymmetric particles would have been copiously produced at the start of our Universe in the Big Bang. Initially these particles would have been in thermal equilibrium. After the temperature of the Universe dropped below the neutralino mass, the neutralino number density would have decreased exponentially. Eventually the expansion rate of the Universe would have overcome the neutralino annihilation rate, resulting in a neutralino density in our Universe today similar to the cosmic microwave background. 

These relic neutralinos could then accumulate in massive celestial bodies in our Universe like our Sun through weak interactions with normal matter and gravity. Over time the neutralino density in the core of the object would increase considerably, thereby increasing the local neutralino annihilation probability. In the annihilation process new particles would be created, amongst which neutrinos. This neutrino flux could be detectable as a localised emission with earth-based neutrino telescopes like ANTARES.

This paper gives a brief overview of the prospects for the detection of neutrinos originating from neutralino annihilation in the Sun with the ANTARES neutrino telescope.

\begin{figure}[b]
\center{
\includegraphics[width=0.45\textwidth,angle=0]{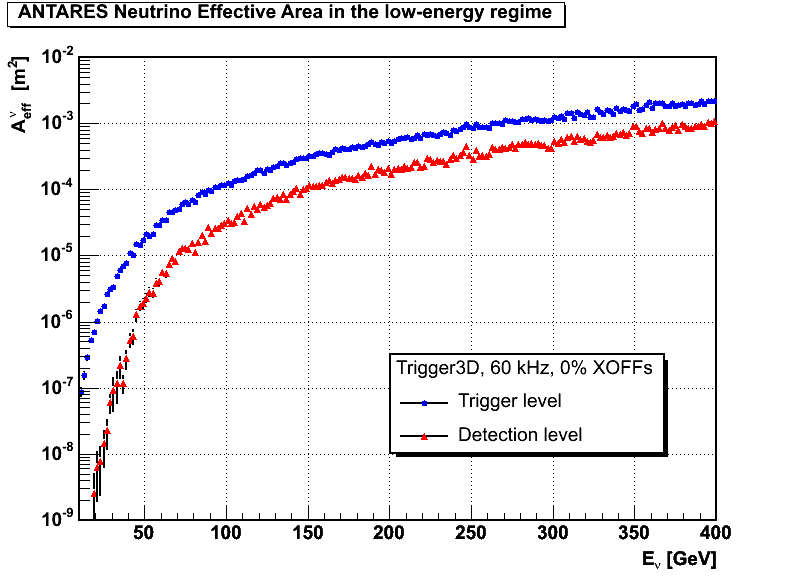}
\caption{The ANTARES Neutrino Effective Area vs. $E_\nu$.}
\label{fig:1}       
}
\end{figure}

\begin{figure*}[t]
\center{
\includegraphics[width=0.8\textwidth,angle=0]{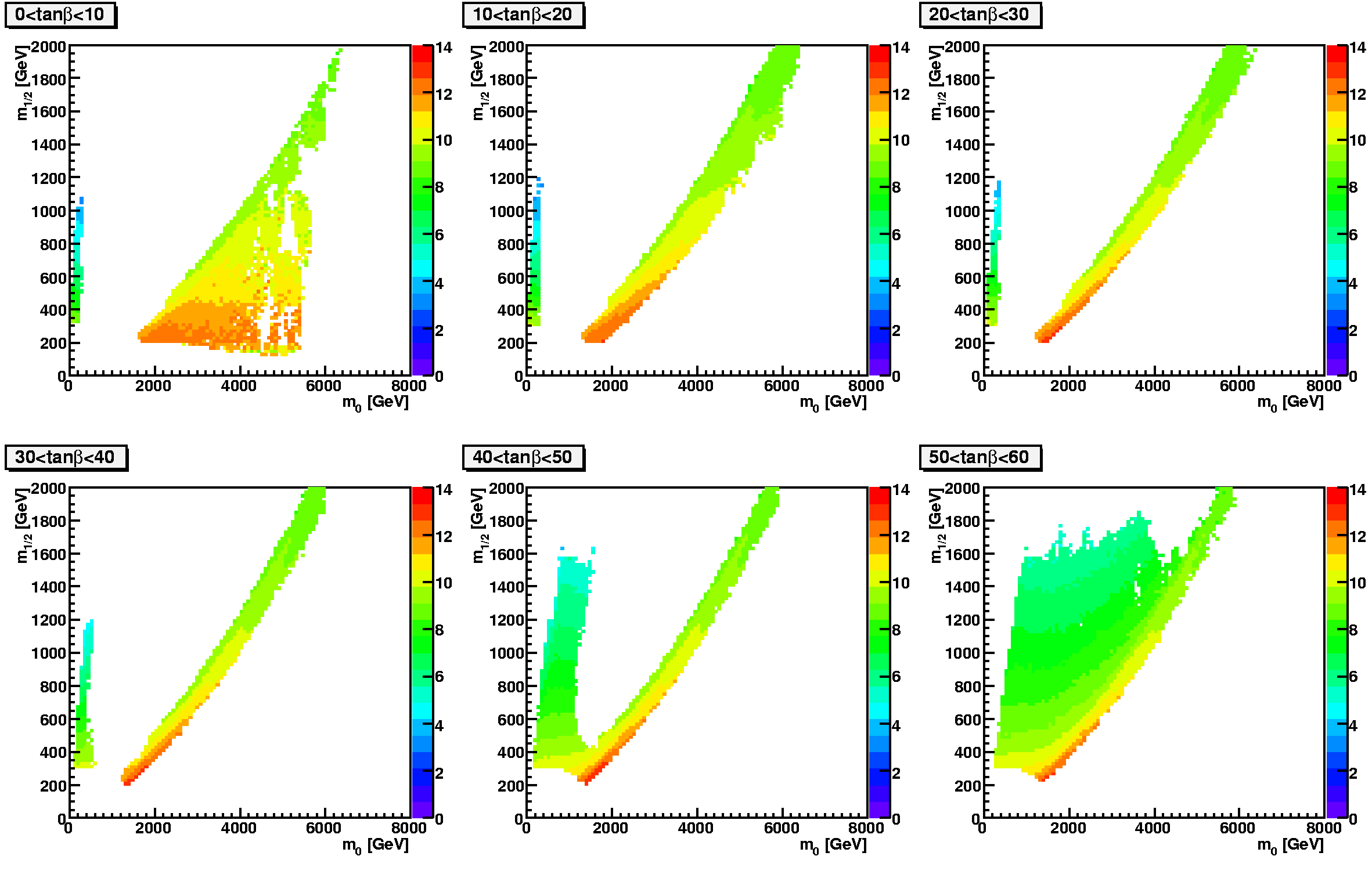}
\caption{Predicted $\nu_\mu+\bar{\nu}_\mu$ flux from the Sun in mSUGRA parameter space.}
\label{fig:2}       
}
\end{figure*}

\section{The ANTARES neutrino telescope}
\label{sec:1} 
The ANTARES undersea neutrino telescope consists of a 3D~grid of 900~photomultiplier tubes arranged in 12~strings, at a depth of 2475~m in the Mediterranean Sea. Three quarters of the telescope have been deployed and half of the detector is already fully operational, making ANTARES the largest neutrino telescope on the northern hemisphere. The angular resolution of the telescope is of the order of one degree at low energy, relevant to dark matter searches, and reaches 0.3 degree at high energies ($>$~10~TeV).

The sensitivity of a neutrino detector is conventionally expressed as the Neutrino Effective Area, $A_{\rm eff}^{\nu}$. The $A_{\rm eff}^{\nu}$ is a function of neutrino energy $E_\nu$ and direction $\Omega$, and is defined as 
\begin{equation}
A_{\rm eff}^{\nu}(E_\nu,\Omega) \;=\; V_{\rm eff}(E_\nu,\Omega)\;\sigma(E_\nu)\;\rho N_A\;P_E(E_\nu,\Omega)
\label{eq:1}
\end{equation}

\noindent where $\sigma(E_\nu)$ is the neutrino interaction cross section, $\rho\,N_A$ is the nucleon density in/near the detector,\linebreak $P_E(E_\nu,\Omega)$ is the neutrino transmission probability\linebreak through the Earth and $V_{\rm eff}(E_\nu,\Omega)$ represents the effective detector volume. This last quantity depends not only on the detector geometry and instrumentation, but is also on the efficiency of the trigger and reconstruction algorithms that are used. 

The ANTARES $A_{\rm eff}^{\nu}$ for upgoing $\nu_\mu$ and $\bar{\nu}_\mu$'s, integrated over all directions as a function of the neutrino energy is shown in Fig.~\ref{fig:1}. The curves represent the $A_{\rm eff}^{\nu}$ after triggering only (``{\em Trigger level}'', in blue) and after reconstruction and selection (``{\em Detection level}'', in red). The increase of the $A_{\rm eff}^{\nu}$ with neutrino energy is mainly due to the fact that $\sigma(E_\nu)$ as well as the muon range are both proportional to the neutrino energy.

The detection rate $R(t)$ for a certain neutrino flux $\Phi(E_\nu,\Omega,t)$ is then defined as

\begin{equation}
R(t) \;=\; \iint\,A_{\rm eff}^{\nu}(E_\nu,\Omega)\;\frac{d\Phi(E_\nu,\Omega,t)}{dE_\nu\,d\Omega}\;dE_\nu\,d\Omega
\label{eq:2}
\end{equation}

\section{Neutralino annihilation in the Sun}
\label{sec:2} 
We calculated the $\nu_\mu+\bar{\nu}_\mu$ flux resulting from neutralino annihilation in the centre of the Sun using the DarkSUSY simulation package \cite{DarkSUSY}. Furthermore, the effects of neutrino oscillations in matter and vacuum as well as absorption were taken into account. The top quark mass was set to 172.5~GeV and the NFW-model for the Dark Matter halo with a local halo density \mbox{$\rho_0 = 0.3$~GeV/cm$^3$} was used. Instead of the general Supersymmetric framework, we used the more constrained approach of minimal Supergravity (mSUGRA). In mSUGRA, models are characterized by four parameters and a sign: A common gaugino mass $m_{1/2}$, scalar mass $m_0$ and tri-linear scalar coupling $A_0$ at the GUT scale ($10^{16}$ GeV), the ratio of vacuum expectation values of the two Higgs fields $tan(\beta)$ and the sign of the Higgsino mixing parameter $\mu$. We considered only $\mu=+1$ models within the following parameter ranges: \mbox{$0<m_0<8000$~GeV,} \mbox{$0<m_{1/2}<2000$~GeV,}\linebreak \mbox{$-3m_0<A_0<3m_0$} and \mbox{$0<tan(\beta)<60$.} The SUSY parameters were subsequently calculated using the\linebreak ISASUGRA package \cite{Isasugra}.

\begin{figure}[b]
  \includegraphics[width=0.45\textwidth,angle=0]{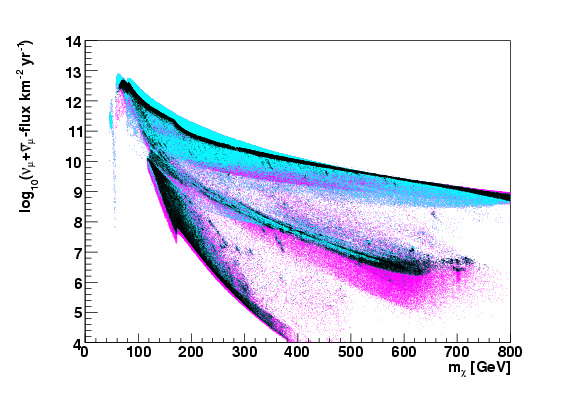}
\caption{Predicted $\nu_\mu+\bar{\nu}_\mu$ flux from the Sun vs. $m_\chi$.}
\label{fig:3}       
\end{figure}

\begin{figure*}[t]
\center{
\includegraphics[width=0.8\textwidth,angle=0]{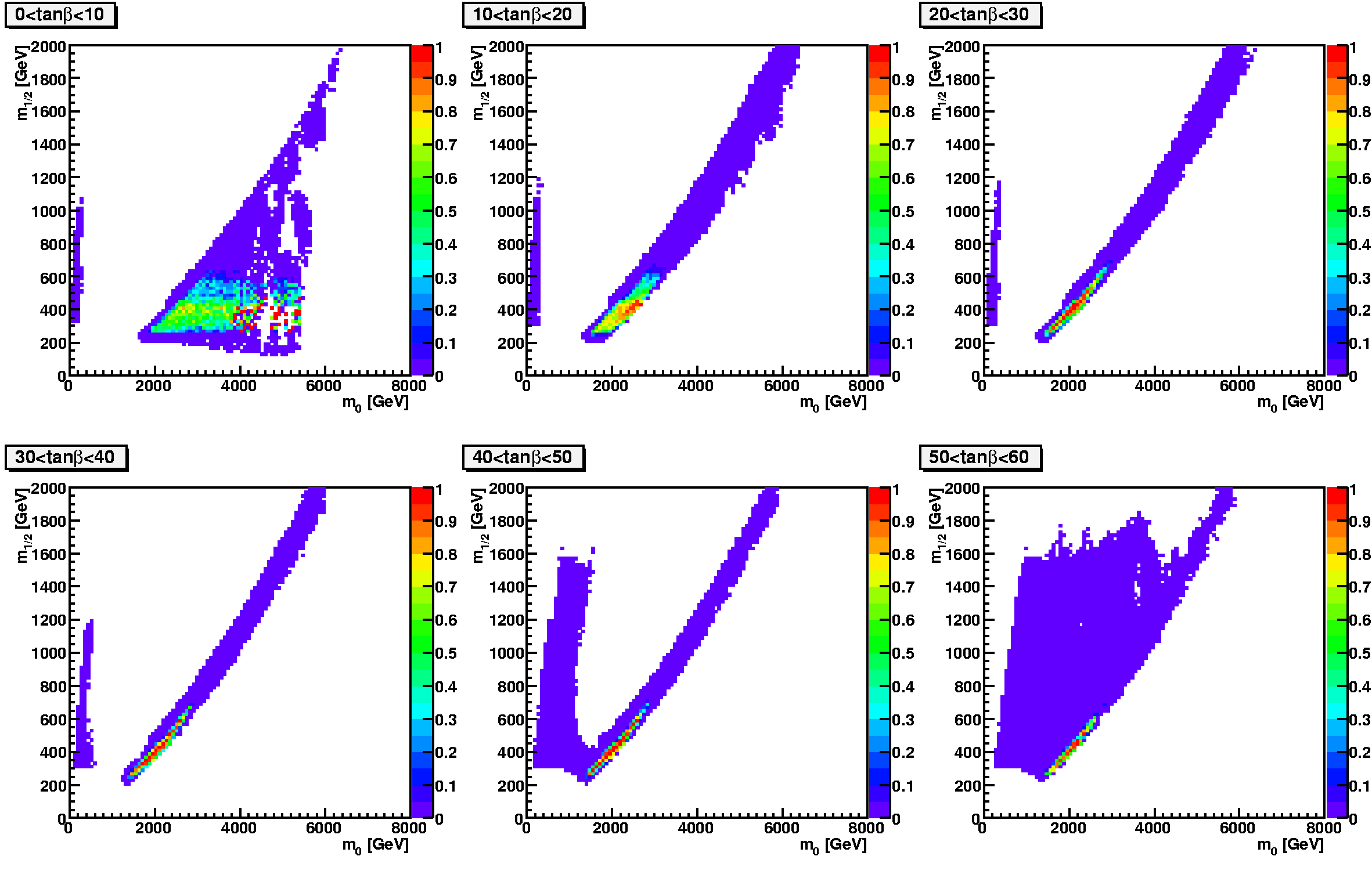}
\caption{mSUGRA models 90\% CL excludable by ANTARES in mSUGRA parameter space.}
\label{fig:4}       
}
\end{figure*}

\pagebreak Only a small subset of all mSUGRA models possess a relic neutralino density $\Omega_\chi h^2$ that is compatible with the Cold Dark Matter energy density $\Omega_{\rm CDM} h^2$ as measured by WMAP \cite{WMAP}. To investigate specifically those mSUGRA models, we sampled the mSUGRA parameter space using a random walk method based on the Metropolis algorithm where $\Omega_\chi h^2$ acted as a guidance parameter \cite{MarkovChain}. 

The resulting $\nu_\mu+\bar{\nu}_\mu$ flux from the Sun per~km$^{2}$ per year, integrated above \mbox{$E_\nu=10$~GeV}, can be seen in the \mbox{$m_0$-$m_{1/2}$~plane} for different ranges of $tan(\beta)$ in Fig.~\ref{fig:2}. The white regions correspond to mSUGRA models without radiative electroweak symmetry breaking, models with $\Omega_\chi h^2>1$, models that are already experimentally excluded, or models where the neutralino is not the lightest superpartner. Models in the so-called ``Focus Point'' region\,\footnote{The region of mSUGRA parameter space around $(m_0,m_{1/2}) = (2000,400)$.} produce the highest neutrino flux: In this region the neutralino has a relatively large Higgsino component \cite{Nerzi}. This enhances the neutralino capture rate through $Z$-boson exchange as well as the neutralino annihilation through the\linebreak \mbox{$\chi\chi\rightarrow WW/ZZ$} channel, resulting in a large flux of relatively high energetic neutrinos.

The $\nu_\mu+\bar{\nu}_\mu$ flux can also be plotted against the neutralino mass $m_\chi$, as is shown in Fig.~\ref{fig:3}. In this plot, the mSUGRA models are subdivided into three categories according to how well their $\Omega_\chi h^2$ agrees with $\Omega_{\rm CDM} h^2$ as measured by WMAP\,\footnote{WMAP: $\Omega_{\rm CDM} h^2 = 0.1126_{-0.013}^{+0.008}$}: \mbox{$\Omega_\chi h^2-\Omega_{\rm CDM}h^2<2\sigma$} (black), \mbox{$0< \Omega_\chi h^2 < \Omega_{\rm CDM} h^2$} (blue) and \mbox{$\Omega_{\rm CDM} h^2 < \Omega_\chi h^2 < 1$} (magenta).

\begin{figure}[b]
  \includegraphics[width=0.45\textwidth,angle=0]{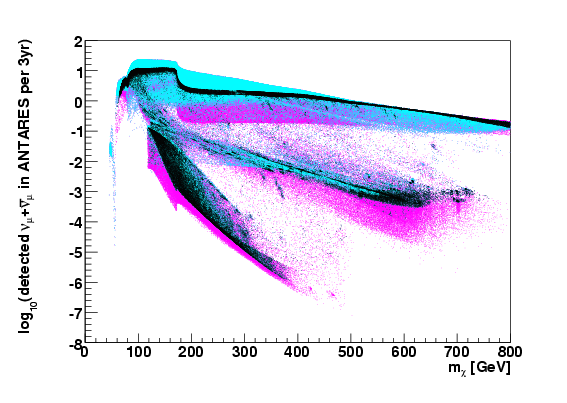}
\caption{ANTARES detection rate per 3~years vs. $m_\chi$.}
\label{fig:5}       
\end{figure}

\section{ANTARES prospects to detect neutralino annihilation in the Sun}
\label{sec:3} 
The ANTARES detection rate (See Eq.~\ref{eq:2}) for the detection of neutralino annihilation in the Sun was calculated as follows: For each mSUGRA model considered in Sect.~\ref{sec:2}, the differential $\nu_\mu+\bar{\nu}_\mu$ flux from the Sun was convoluted with the Sun's zenith angle distribution as well as the ANTARES $A_{\rm eff}^{\nu}$ (see Eq.~\ref{eq:1} and Fig.~\ref{fig:1}). The resulting detection rate in ANTARES per 3~years is shown as a function of the neutralino mass in Fig.~\ref{fig:5}. The color coding in the plot corresponds to the one used in Fig.~\ref{fig:3}.

\begin{figure}[t]
  \includegraphics[width=0.45\textwidth,angle=0]{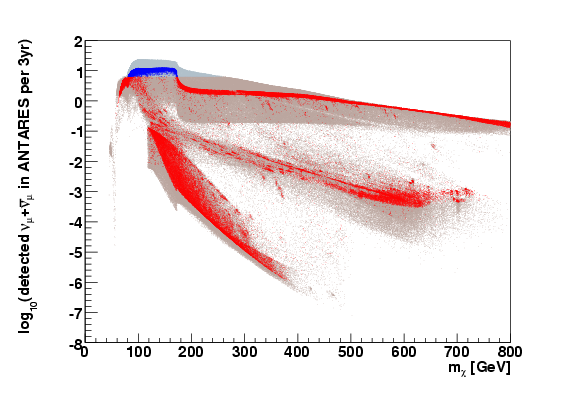}
\caption{mSUGRA models 90\% CL excludable by ANTARES per 3~years vs. $m_\chi$.}
\label{fig:6}       
\end{figure}

The ANTARES exclusion limit for the detection of neutralino annihilation in the Sun was calculated as follows: As can be seen from Fig.~\ref{fig:5}, the expected detection rates for all mSUGRA model considered in Sect.~\ref{sec:2} are small. Therefore the Feldman Cousins approach \cite{FeldmanCousins} was used to calculate 90\%~CL exclusion limits. The two sources of background were taken into account as follows: Since we know the Sun's position in the sky, the atmospheric neutrino background (Volkova parametrisation) was considered only in a 3~degree radius search cone around the Sun's position. After applying the event selection criteria used to determine the $A_{\rm eff}^{\nu}$ in Fig.~\ref{fig:1}, the misreconstructed atmospheric muon background was considered to be 10\% of the atmospheric neutrino background. mSUGRA models that are excludable at 90\%~CL by ANTARES in 3~years are shown in blue in Fig.~\ref{fig:6}, those that are non-excludable are shown in red. Bright colors indicate models which have \mbox{$\Omega_\chi h^2-\Omega_{\rm CDM}h^2<2\sigma$}. The fraction of ANTARES excludable models in mSUGRA parameter space is shown in Fig.~\ref{fig:4}.

\begin{figure}[b]
  \includegraphics[width=0.45\textwidth,angle=0]{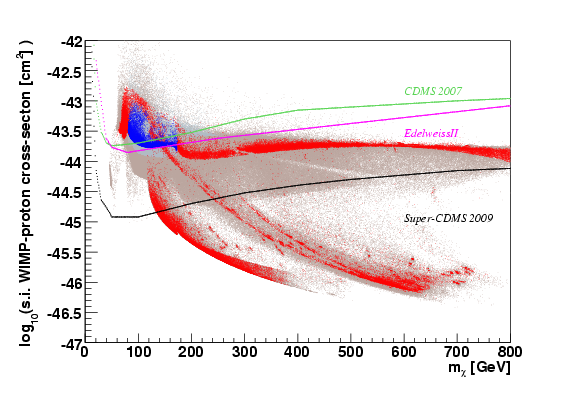}
\caption{Spin-independent $\chi p$~cross section vs. $m_\chi$.}
\label{fig:7}       
\end{figure}

\begin{figure}[t]
  \includegraphics[width=0.44\textwidth,angle=0]{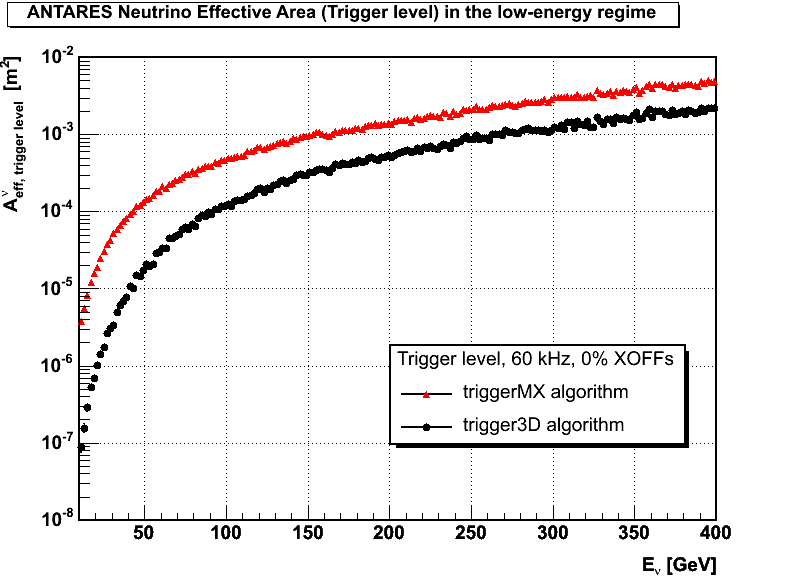}
\caption{The ANTARES Neutrino Effective Area at the trigger level vs. $E_\nu$.}
\label{fig:8}       
\end{figure}

\section{Comparison to direct detection}
To compare with direct detection experiments, the spin-independent $\chi p$~cross section versus the neutralino mass for all mSUGRA models considered in Sect.~\ref{sec:2} is shown in Fig.~\ref{fig:7}. The color coding in the plot corresponds to the one used in Fig.~\ref{fig:6}. The limits in this plot were taken from the Dark Matter Limit Plot Generator \cite{DirectDetection}. The spin-independent cross section is driven by CP-even Higgs boson exchange \cite{Nerzi}. Therefore, mSUGRA models in which the neutralino is of the mixed gaugino-Higgsino type will produce the largest cross sections. This implies a correlation between the models that are excludable by direct detection experiments and models excludable ANTARES, as can be seen from Fig.~\ref{fig:7}.

\section{Conclusion \& Outlook}
Nearly half of the ANTARES detector has been operational since January this year. The detector is foreseen to be completed in early 2008. The data show that the detector is working within the design specifications. 

As can be seen from Fig.~\ref{fig:4}, mSUGRA models that are excludable by ANTARES at 90\%~CL are found in the Focus Point region. The same models should also be excludable by future direct detection experiments, as is shown in Fig.~\ref{fig:7}.

To improve the ANTARES sensitivity, a directional trigger algorithm has recently been implemented in the data acquisition system. In this algorithm, the known position of the potential neutrino source is used to lower the trigger condition. This increases the trigger efficiency, resulting in a larger $A_{\rm eff}^{\nu}$. In Fig.~\ref{fig:8}, the $A_{\rm eff}^{\nu}$ at the trigger level for the standard- and the directional trigger algorithm are shown in black (``{\em trigger3D}'') and red (``{\em triggerMX}'') respectively.


\end{document}